\documentclass[reprint,superscriptaddress,amsmath,amssymb,aps,pra]{revtex4-1}

\usepackage{standalone}
\usepackage[dvipsnames]{xcolor}
\usepackage{graphicx}
\usepackage{dcolumn}
\usepackage{bm}
\usepackage[colorlinks]{hyperref}
\usepackage{pgfplots}

\pgfplotsset{compat=newest}
\usepackage[utf8]{inputenc}
\usepackage{pgfplots}
\usepackage{physics}
\usepackage{amsmath}
\usepackage{braket}
\usepackage{filecontents}

\usepackage{soul}

\usepackage[colorlinks]{hyperref}
\newcommand\myshade{80}
\colorlet{mylinkcolor}{ForestGreen}
\colorlet{mycitecolor}{Red}
\colorlet{myurlcolor}{violet}
\usepackage[dvipsnames]{xcolor}
\usepackage{changes}

\hypersetup{
  linkcolor  = mylinkcolor!\myshade!black,
  citecolor  = mycitecolor!\myshade!black,
  urlcolor   = myurlcolor!\myshade!black,
  colorlinks = true
}

\newcommand{\UNIBO}{
Dipartimento di Fisica e Astronomia, Universit{\`a} di Bologna, \\
Via Berti-Pichat 6/2, I-40126 Bologna, Italy}

\newcommand{\LPTWON}{
Univ. Bordeaux, CNRS, IOGS, LP2N, UMR 5298, F-33400 Talence, France}

\begin{document}

\preprint{APS/123-QED}

\title{Fast control of atom-light interaction in a narrow linewidth cavity}

\author{A. Bertoldi}
\email{andrea.bertoldi@institutoptique.fr}
\affiliation{\LPTWON}

\author{C.-H. Feng}
\affiliation{\LPTWON}

\author{D. S. Naik}
\affiliation{\LPTWON}

\author{B. Canuel}
\affiliation{\LPTWON}

\author{P. Bouyer}
\affiliation{\LPTWON}

\author{M. Prevedelli}
\affiliation{\UNIBO}

\date{\today}

\begin{abstract}

We propose a method to exploit high finesse optical resonators for light assisted coherent manipulation of atomic ensembles, overcoming the limit imposed by the finite response time of the cavity. The key element of our scheme is to rapidly switch the interaction between the atoms and the cavity field with an auxiliary control process as, for example, the light shift induced by an optical beam. The scheme is applicable to many different atomic species, both in trapped and free fall configurations, and can be adopted to control the internal and/or external atomic degrees of freedom. Our method will open new possibilities in cavity-aided atom interferometry and in the preparation of highly non-classical atomic states.

\end{abstract}

\pacs{Valid PACS appear here}

\maketitle

Narrow linewidth cavities are key devices in fundamental physics \cite{Abbott2016}, metrology \cite{Robinson2019}, and they underpin the incessant progress in the study of light-matter interaction \cite{Ritsch2013,Haslinger2017}.
In atom interferometry (AI) high finesse optical cavities can improve the instrument sensitivity by allowing very high momentum transfer beamsplitters \cite{Hamilton2015,Riou2017,Nourshargh-2-2020,Jaffe2018,Kristensen2020}. Cavities with long length $L$ are instead sought for gravitational wave (GW) detection \cite{Canuel2018} to increase the strain sensitivity proportionally to $L$. The combination of high finesse $\cal F$ and long $L$ has the effect of reducing its linewidth $\Delta \nu = c /\left( 2 n L {\cal F} \right)$, where $c$ is the speed of light in vacuum, and $n$ the index of refraction inside the cavity. Despite the promise of increased sensor performance, it has been pointed out in \cite{Dovalelvarez2017} that a limitation exists for $\Delta \nu$, beyond which the pulses used to coherently manipulate the atomic wavefunction undergo important deformation, and where the effective optical power enhancement worsens. In short, the inherent frequency response of the cavity sets a physical limit to the product $L \cal F$, and forbids adopting narrow linewidth resonators for manipulating matter waves.

\begin{figure}[b]
  \includegraphics[scale=1.2]{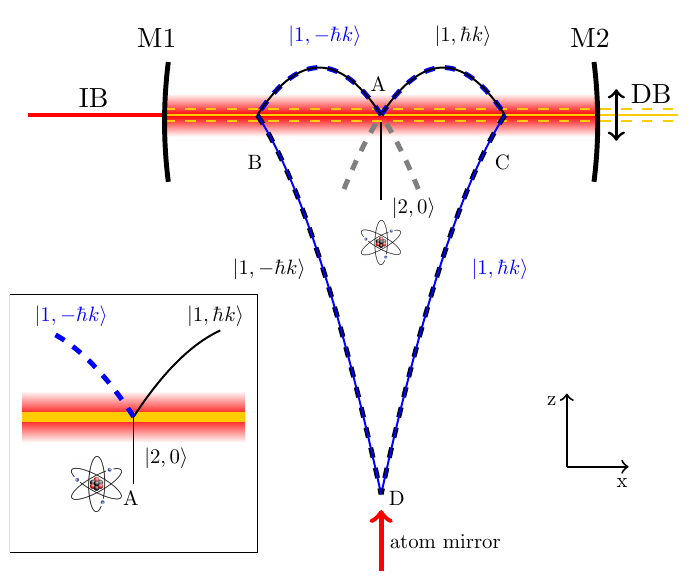}
  \caption{Schematic of the proposed experimental setup not to scale: the atomic ensemble, initially in the state $\ket{2}$ and moving in the $z$ direction, crosses the cavity-enhanced \textit{IB}, and is split in the region A (see inset) in two paths with opposite horizontal velocity $\pm v_r$. The two parts of the wavefunction are horizontally reflected with a mirror pulse in the regions B and C; in D their vertical velocity is inverted, and after a second mirror pulse, again in C and B, they are recombined in A with a last split pulse. The two trajectories at the output of the interferometer are shown in gray. The horizontal \textit{DB} (yellow), not resonant with the cavity, is shone on the atoms and vertically follows their motion to have an optimal overlap. M1, M2: cavity mirrors; $v_r$, \textit{DB} and the atom labelling are defined in the main text.}
  \label{fig:gravimeter}
\end{figure}

In this Letter, we propose a novel scheme to coherently manipulate the atomic wavefunction in a narrow linewidth cavity, where the interaction is pulsed not by changing the intensity of the intracavity standing wave, but by modulating the coupling between the intracavity light and the atoms, using an auxiliary process. The cavity enhanced laser is always injected in the optical resonator, hence its intensity is constant in time. The main approach we analyze exploits light-shift engineering of the atomic levels, a technique adopted in several contexts concerning cold atoms, e.g. to cancel the trapping light perturbation in optical lattice clocks \cite{Katori2003}, laser cool atoms to BEC \cite{Stellmer2013}, and precisely characterize the geometry of an optical cavity \cite{Bertoldi2010}.

\begin{figure}[t!]
  \includegraphics[scale=1.1]{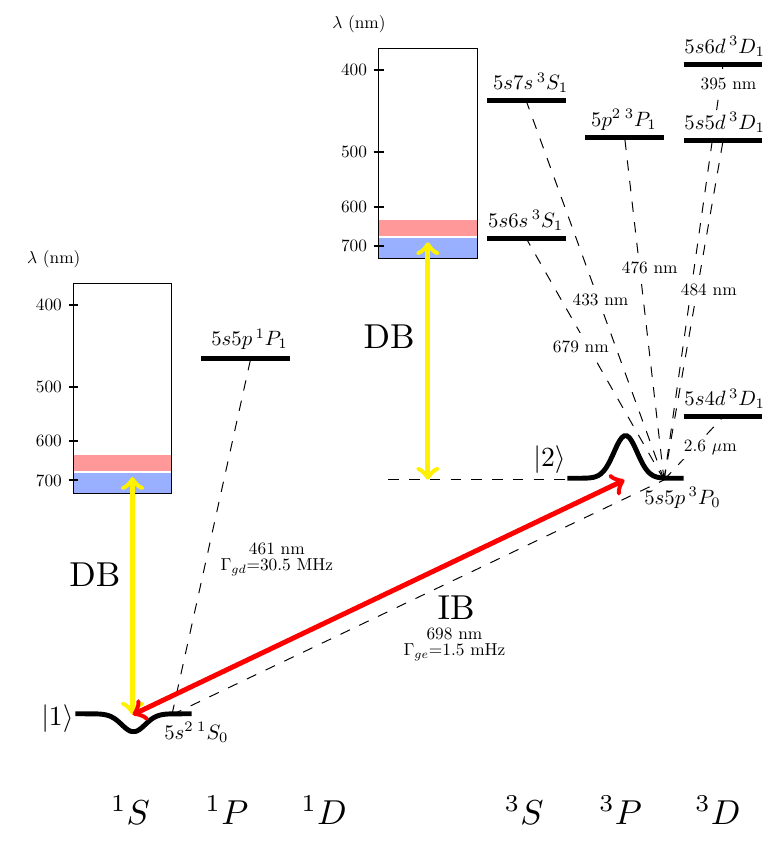}
  \caption{ Diagram with the relevant levels for $^{87}\mathrm{Sr}$ atoms. The red arrow shows the \textit{IB} resonant to the $\ket{1} \rightarrow \ket{2}$ transition at 698 nm adopted for the coherent manipulation of matter waves; the yellow arrows mark the \textit{DB} used to shift the two levels $\ket{1}$ and $\ket{2}$. The action of the \textit{DB} is considered when varying its wavelength over the range [380--740] nm, indicated by the vertical bars referenced to the two levels $\ket{1}$ and $\ket{2}$. The narrow red (blue) bands indicate the spectral interval where the \textit{DB} with parallel (perpendicular) polarization constitutes an effective switch for the coherent action of \textit{IB}, by light shifting in a differential fashion the clock levels $\ket{1}$,$\ket{2}$. The bands have been obtained as defined in Fig. \ref{fig:LSvsScatt}. The level structure has been taken from \cite{nist,LudlowPhD}.}
  \label{fig:scheme}
\end{figure}

For the sake of clarity, we focus our study on the example of an AI-based gravimeter using $^{87}\mathrm{Sr}$ atoms driven on the clock transition and vertically launched in free fall to cross a horizontal cavity (see Fig. \ref{fig:gravimeter}). Our scheme can easily be extended to other configurations relying on cavity-enhanced light-pulses to manipulate the atomic state. The coherent manipulations are performed at each passage of the atoms in the cavity, by pulses shorter than the transit interval in the \textit{Interferometric Beam} (IB). We consider a 4 pulse sequence interferometer \cite{Marzlin1996} based on the double-diffraction scheme \cite{Lvque2009} and with a time separation $T-2T-T$ ($T$=0.25 s will be assumed later for numerical evaluations). This geometry relies on a single horizontal \textit{IB} thanks to a vertical reflection at the middle of the sequence, and other similar configurations could be considered \cite{Schubert2019}. At $t$=0 the atoms are in $A$ (Fig.~\ref{fig:gravimeter}), in state $|2\rangle$ with a velocity $(v_x,v_z)=(0,-gT/2)$, where $g$ is the local acceleration of gravity. Here, they interact with a first splitting pulse. The atomic wavefunction is divided in two components in state $|1\rangle$ with opposite horizontal velocities $v_x=\pm v_r$, where $v_r=\hbar k/m$ is the recoil velocity. At $t$=$T$/2 the atoms reach the apogees of the trajectories at a height of $-gT^2/8$ above the \textit{IB}. At $t=T$ the atoms are again in the \textit{IB} where a mirror pulse reverses their respective horizontal velocity without changing their internal state. At $t=2T$ the two parts of the wavefunction cross in $D$ with a vertical velocity $v_z=3gT/2$; they are vertically reflected by optical means \cite{Dickerson2013} and complete the second, symmetric half of the interferometric sequence. Note that during the free evolution the atoms are always in the same internal state, thus cancelling many systematic effects, and that the resonance condition for the \textit{IB} is the same for all the pulses.

The narrow linewidth cavity is locked to the linearly polarized \textit{IB}; the beam intensity is thus increased and the spatial mode filtered. The intracavity enhanced intensity of \textit{IB} is chosen to have a Rabi frequency $\Omega_{\mathrm{R}}$ of $2\pi \times$5 kHz. For a double diffraction, the split and mirror pulses must have a length of $\tau_s=\pi/(\sqrt{2} \, \Omega_{\mathrm{R}})$ and $\tau_m=2\tau_s$ respectively \cite{Lvque2009} so we obtain $\tau_s\simeq 70 \mu\mathrm{s}$. The pulses do not couple to spurious momentum states as long as $\tau_m \omega_r \gg 1$ where $\omega_r$ is the recoil frequency i.e. about 59 kHz for $^{87}\mathrm{Sr}$.

As mentioned above, we focus on alkali--earth atoms, more specifically $^{87}\mathrm{Sr}$, where \textit{\textit{IB}} is tuned on the narrow transition at 698 nm defined by the levels $\ket{1} \equiv \, ^1$S$_0$ and $\ket{2} \equiv \, ^3$P$_0$ \cite{Campbell2017} (see Fig. \ref{fig:scheme}), to implement the coherent manipulation scheme proposed in \cite{Graham2013} and recently demonstrated in \cite{Hu2017} for $^{88}\mathrm{Sr}$. An additional \textit{Dressing Beam (DB)} differentially shifts the levels $\ket{1}$ and $\ket{2}$, breaking the resonance condition for \textit{IB}. Modulating the intensity of \textit{DB} will allow to switch the resonance with \textit{IB} on and off \footnote{Other kinds of atoms and manipulation schemes can be considered. For example, in the case of Raman transitions on alkali atoms, one could use, as \textit{DB}, a laser on the D2 transition, which creates a light shift with opposite signs on the two hyperfine levels connected by the Raman transition.}.

For numerical application, we consider a narrow linewidth cavity that can fit in a conventional laboratory: the cavity parameters are set to be L = 2 m and ${\cal F}$=10$^5$ ($\Delta \nu = 750 \, \rm{Hz}$). In this configuration, the fast amplitude modulation of the \textit{IB} to implement the interferometric pulses would generate strong pulse deformations in the cavity which is detrimental to the sought power enhancement \cite{Dovalelvarez2017}.

The $^{87}$Sr atoms are considered at very low temperature, prepared in the $\ket{2}$ state and launched vertically, to reach point A. The spatial extension of the atomic cloud is assumed $<100 \, \mu\rm{m}$ during all the duration of the interferometric sequence. This will require adopting delta-kick collimation techniques \cite{Muntinga2013,Kovachy2015} to prepare the atomic source. The cavity waist is set to be $1 \, \rm{mm}$, so as to obtain a rather homogeneous manipulation of the atomic ensemble on its axis, even when taking into account the vertical displacement of the cloud during the manipulation. To obtain the required $\Omega_{\mathrm{R}}$, the intracavity power of the \textit{IB} must be $\mathrm{P}_0 \simeq 286\,\mathrm{mW}$ \cite{Loriani2019}, which means an input power $\mathrm{P_{in}} \simeq 390\,\mu\mathrm{W}$, if two lossless mirrors with equal reflectivity are considered for the cavity. 

The atomic interaction with the \textit{IB} is controlled with the \textit{DB} (in yellow in Fig. \ref{fig:gravimeter} and Fig. \ref{fig:scheme}), whose role is to induce an additional energy shift $\Delta \omega_{21}$ on the $\ket{1} \rightarrow\ket{2}$ transition, so as to remove the resonance condition for the cavity enhanced \textit{IB}. This solution has also the technical advantage of avoiding to have to re-lock to the cavity the laser generating the \textit{IB} at each pulse. To calculate $\Delta \omega_{21}$ when varying the \textit{DB} wavelength over the range $380 \, \rm{nm} < \lambda < 740 \, \rm{nm}$ we considered the relevant levels shown in Fig. \ref{fig:scheme}, and the transition parameters reported in \cite{LudlowPhD,nist}.

A single \textit{DB} along the cavity axis (see Fig. \ref{fig:gravimeter}) can dress the atoms along both interferometric trajectories during their passages in the \textit{IB}; the cavity mirrors must be transparent at the \textit{DB} wavelength, to maintain a high bandwidth for the variation of the beam intensity, and to allow its vertical translation to track the atomic motion as described below. A bias magnetic field $B$ is added in the vertical direction to define the  quantization axis. The \textit{DB} is linearly polarized either parallel or perpendicular to $B$.

\begin{figure}[t!]
  \includegraphics[width=.5\textwidth]{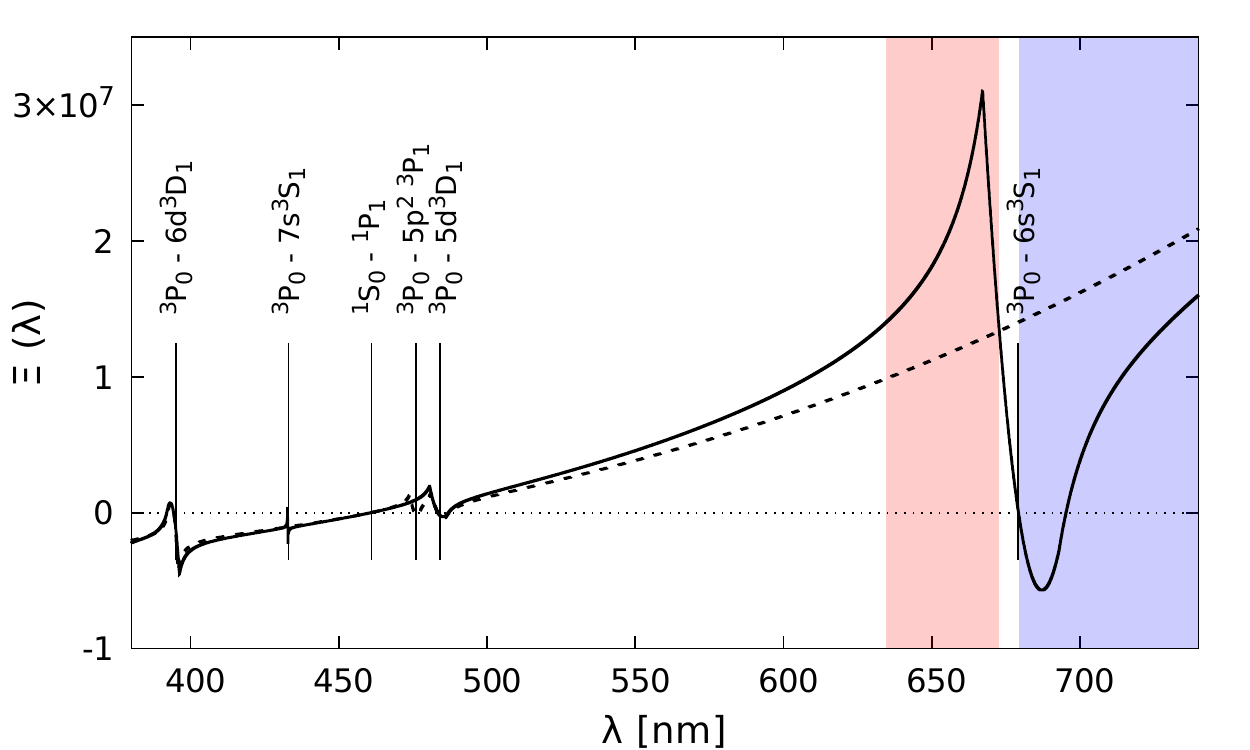}
  \caption{\label{fig:LSvsScatt} Ratio $\Xi(\lambda)$ between the light shift induced on the $\ket{1} - \ket{2}$ transition and the overall scattering rate by a laser at a wavelength $\lambda$. The solid (dashed) curve refers to the \textit{DB} polarization parallel (perpendicular) to the magnetic field. The regions where $\Xi(\lambda) > 1.4 \times 10 ^7$ are indicated with a red (blue) vertical band for parallel (perpendicular) polarization of the \textit{DB}. The wavelengths of the relevant transitions contributing to the atomic polarizability in the visible spectrum are indicated with vertical lines and labelled.}
\end{figure}

The main unwanted side effect of the \textit{DB} on the atomic system is the scattering of photons at a rate $\Gamma_{\rm{sc}}$ on the two levels of interest, which represents a decoherence channel. Other effects, like the DB wavefront aberrations, are not considered here; their impact, however, is highly reduced in the differential configuration provided by a gravity-gradiometer. By dividing $\Delta \omega_{21}$ by $\Gamma_{\rm{sc}}$ we obtain a normalized light shift $\Xi (\lambda)$, plotted in Fig. \ref{fig:LSvsScatt} for a \textit{DB} polarized along the magnetic field (continuous curve) and orthogonal to it (dashed curve).

To optimize the \textit{DB} parameters, we start by arbitrarily fixing the maximum probability to scatter a photon from the \textit{DB} during the whole interferometric sequence to 3\%, which means a subsequent reduction of the interferometer contrast of the same order. Considering the atomic vertical speed at each passage in the cavity ($-gT/2 \sim$1.25 m/s with our choice of $T$) it means a maximum nominal scattering rate $\Gamma_{\rm{sc}} \sim$1 Hz at the center of the \textit{DB} \footnote{The value is obtained by considering the \textit{DB} always centered on the atomic cloud during the 4 crossings of the \textit{IB} and over the whole vertical spatial interval clipped by the cavity, i.e. 10 times its waist.}.

The second parameter to set is the minimum differential light shift required to effectively suppress the Rabi oscillation between states $\ket{1}$ and $\ket{2}$. To this aim, the generalized Rabi frequency $\tilde{\Omega}_R= \left( \Omega_R^2+\Delta\omega_{12}^2 \right)^{1/2}$ when the \textit{DB} is on must be $\gg \Omega_{\textrm{R}}$, and the rms uncertainty of the interferometric phase due to the residual Rabi oscillation is equal to:
\begin{equation}
  \delta \phi = \frac{\Omega_{\textrm{R}}}{\sqrt{2} \; \Delta \omega_{21}} , \nonumber
  \label{eq:conditionRabi}
\end{equation}
if $\Delta \omega_{21} \gg \Omega_{\textrm{R}}$ \footnote{The phase uncertainty is calculated for a balanced atomic state; a state unbalance $\delta \equiv N_1-N_2$ determines (i) a reduction factor $\sqrt{1-\delta^2}$ on the phase uncertainty, and (ii) a systematic phase shift equal to $\sqrt{1-\delta^2} \cdot \Omega/\Delta$}. We set a threshold of 3$\times10^{-3}$ - i.e. the QPN of 10$^5$ atoms - for the overall phase uncertainty due to the residual Rabi oscillation during the 4 atomic passages in the cavity. Any coherent evolution other than between states $\ket{1}$ and $\ket{2}$ (see Fig. \ref{fig:scheme}) has been neglected in this calculation. This assumption is valid whenever the \textit{DB} is far from the specific transition frequencies.

To simultaneously satisfy the requirements on the scattering rate and residual Rabi oscillation, one must have $| \Xi(\lambda) | > 1.4 \times 10^7$. In the visible this condition is satisfied for a linearly polarized \textit{DB} along (perpendicular to) the bias magnetic field \textit{B} for $633 \, \rm{nm} < \lambda < 672 \, \rm{nm}$ ($\lambda > 679 \, \rm{nm}$), as shown by the colored bands in Fig. \ref{fig:LSvsScatt}. At $\lambda=672 \, \rm{nm}$, for example, a \textit{DB} with a waist of 100 $\mu$m and power $\sim 10\,\mathrm{W}$ determines a residual oscillation amplitude below the threshold mentioned above for a scattering probability $<2$\%. To avoid increasing the required power, the last mirror directing the beam on the atoms can be mounted on a fast and precise translation stage, in order to track with the DB the atomic cloud's motion in the IB. Other wavelengths also satisfy the above requirements, even in a stricter fashion: in the interval [1350 nm--2.5 $\mu$m] the ratio $| \Xi(\lambda) |$ is compatible with an instrument sensitivity below the QPN of 10$^9$ atoms with a 5\% contrast reduction, and even better parameters are obtained at CO$_2$ laser wavelengths. Nevertheless, the required laser power for these wavelengths is in the kW range.

In the configuration studied previously, the interaction between the atoms and the \textit{IB} is turned off when the \textit{DB} is on. A trade-off must be found between having a large \textit{DB} power to effectively switch off the IB, and minimizing the residual scattering rate it causes. Another scheme, which is not analyzed in details in this publication, consists in using the \textit{DB} to turn on the interaction. Photon scattering is strongly reduced because the \textit{DB} is only on during the coherent manipulation pulses. As a consequence, the \textit{DB} can be set closer to a transition between $\ket{2}$ and an excited level. This has three advantages: (i) lowers the \textit{DB} power; (ii) adds the \textit{IB} detuning as a parameter to reduce even further the residual Rabi oscillation; (iii) makes \textit{CB} unnecessary, with a suitable choice of the \textit{DB}'s wavelength and intensity. The price to pay is that the control of the coherent manipulation now depends not only on the stability of the \textit{IB}, but also on the stability of the \textit{DB}.

Two other effects of the cavity can affect the coherent manipulation. First, the intracavity \textit{IB} light intensity can decay during the time the \textit{DB} is turned off, because of the modified effective atomic index of refraction that shifts the cavity resonance \cite{Long2007}. The cavity narrow bandwidth prevents, however, the intracavity field to evolve significantly during the duration of the light pulses, which is much shorter than the cavity response time. Second, the atomic absorption can spoil the cavity linewidth \cite{Salvi2018}; again, for the adopted parameters, namely the number of atoms and the threshold set on the allowed scattering rate, the effect has been evaluated to be negligible. \medskip

We now focus on the specific case of a cavity-aided gravity-gradiometer for GW detection which motivated this proposal \cite{Dovalelvarez2017}. An instrument with a long baseline length $L$ and a 4 pulse sequence gives a phase sensitivity
\begin{equation}
\delta \phi \approx 8 k L h_+ \sin \left(2 \pi f T \right) \sin^2 \left( \frac{2 \pi f T}{2}\right), \nonumber
\end{equation}
to a plus-polarized GW of frequency $f$ \cite{Graham2013,Canuel2018}. Considering $L$=10 km (i.e. the design value for the Einstein Telescope \cite{Punturo2010}), and a finesse ${\cal F}$=100 (i.e lower than the system design Finesse of aLIGO, which is 450 \cite{Aasi2015}) one obtains $\Delta \nu$=150 Hz. Such value excludes the possibility to realize interferometric pulses shorter than 1 ms by varying the intensity of the \textit{IB} injected in the cavity. Adding a \textit{DB} to design the pulses removes the limitation on the minimum pulse length.

With our parameters and for T=0.25 s, the peak strain sensitivity is $h_+\sim 1.5 \times 10^{-17}/\sqrt{\rm{Hz}}$ at 2 Hz for a shot noise limited detection of $10^5$ atoms per second. Improved strain sensitivities can be obtained by adopting a higher atomic flux, and exploiting the cavity to implement sub-shot-noise sensitivity \cite{Hosten2016,Salvi2018} and large momentum splitting \cite{Chiow2011,Gebbe2019,Pagel2019}. The latter can be achieved by inserting several $\pi$ pulses in the sequence as described in \cite{Lvque2009}, and using the amplitude of the \textit{DB} to maintain the Doppler shift compensation. At the same time background noise signals, arising from the residual phase noise induced by the out-of-resonance \textit{IB}, must be proportionally reduced, exploiting the common mode rejection ratio of the gradiometer or improving the frequency and amplitude stability of the \textit{DB}. The sensitivity curve can be shifted at lower frequency by increasing the atomic interrogation time, which requires to adapt accordingly the specifications of the atom mirror pulse \cite{Schubert2019}.

We have proposed a new coherent manipulation scheme to bypass the limitations of cavity linewidth in cavity-aided AI. Our method enables fast and pulsed manipulation of matter waves with the intracavity resonant light without any restrictions on cavity length and finesse. The scheme described here relies on light-shift engineering to control the atomic coupling on a narrow optical transition to the light stored in the cavity. It could be extended to manipulation schemes with freely-falling or trapped atoms \cite{Chiow2011,Gebbe2019,Pagel2019}, or relying on moderately narrow transitions with relatively higher single-photon Rabi frequency \cite{Rudolph2020}. Other control processes could be adopted, such as magnetic field induced spectroscopy \cite{Taichenachev2006}, three photon resonance \cite{Hong2005}, DC Stark effect \cite{Yatsenko2002}; notably, they could introduce a more homogeneous control of the coherent switching, and a mitigation of the related aberration issue.

This method opens perspectives to push the use atomic cavities in long baseline atom interferometers, such as proposed for GW detection, and to exploit high finesse (narrow linewidth) cavities to improve the spatial filtering of the coherent manipulation beams \cite{Dovalelvarez2017}. This can be used for shorter pulses, large momentum transfer atom optics, and may even lead to universal AI \cite{Cronin2015}. High finesse cavities can also be used for quantum enhanced measurements \cite{Bishof2013,Kohlhaas2015,Cox2016,Hosten2016,Salvi2018}, and could open new avenues for the creation of macroscopic quantum states in optomechanics, by providing a fast and deterministic way to control the transparency of a BEC \cite{Lombardo2015}. \medskip

\textit{Note:} Another scheme to generate pulses beyond the limit set by the cavity bandwidth has been reported recently; it uses intracavity frequency modulation on circulating, spatially resolved pulses \cite{Nourshargh2020}. \medskip

We thank T. Freegarde for a critical reading of our manuscript. This work was partly supported by the ``Agence Nationale pour la Recherche" (grant EOSBECMR \# ANR-18-CE91-0003-01, grant ALCALINF \# ANR-16-CE30-0002-01, and grant MIGA \# ANR-11-EQPX-0028), the European Union (EU) (FET-Open project CRYST$^3$), IdEx Bordeaux - LAPHIA \# ANR-10-IDEX-03-02 (grant OE-TWC) Horizon 2020 QuantERA ERA-NET (grant TAIOL \# ANR-18-QUAN-00L5-02), and Conseil Régional d’Aquitaine (grant IASIG-3D and grant USOFF).


%

\end{document}